\documentclass{svproc}

\usepackage{url}

\usepackage{graphicx}

\usepackage{amsmath}
\DeclareMathOperator{\erf}{erf}

\usepackage{multirow}

\usepackage[misc,geometry]{ifsym}

\usepackage[usenames,dvipsnames]{xcolor}
\definecolor{backgroundcolor}{RGB}{240, 240, 240}
\usepackage{listings,chngcntr}
\begin{document}
%\counterwithin{lstlisting}{section}  % listing caption number format as 1.1, 1.2, 2.1, ...
\renewcommand{\thelstlisting}{\arabic{lstlisting}} % listing caption number format as 1, 2, 3, ...
\mainmatter              % start of a contribution
\title{Black-Scholes Option Pricing on Intel CPUs and GPUs: Implementation on SYCL and Optimization Techniques}
\titlerunning{Black-Scholes Option Pricing on Intel CPUs and GPUs}  % abbreviated title (for running head)
\author{Elena~Panova\inst{1} \and Valentin~Volokitin\inst{1,2} \and
Anton~Gorshkov\inst{1} \and Iosif~Meyerov\inst{1,2(\textrm{\Letter})}}
\authorrunning{Elena~Panova et al.} % abbreviated author list (for running head)
\tocauthor{Elena~Panova, Valentin~Volokitin, Anton~Gorshkov, Iosif~Meyerov}
\institute{oneAPI Center of Excellence, Lobachevsky University, Nizhni Novgorod 603950, Russia
\and
Mathematical Center, Lobachevsky University, Nizhni Novgorod 603950, Russia
%\and
%{\color{red}...}
\email{meerov@vmk.unn.ru}}

\maketitle              % typeset the title of the contribution

\begin{abstract}
The Black-Scholes option pricing problem is one of the widely used financial benchmarks. We explore the possibility of developing a high-performance portable code using the SYCL (Data Parallel C++) programming language. We start from a C++ code parallelized with OpenMP and show optimization techniques that are beneficial on modern Intel Xeon CPUs. Then, we port the code to SYCL and consider important optimization aspects on CPUs and GPUs (device-friendly memory access patterns, relevant data management, employing vector data types). We show that the developed SYCL code is only 10\% inferior to the optimized C++ code when running on CPUs while achieving reasonable performance on Intel GPUs. We hope that our experience of developing and optimizing the code on SYCL can be useful to other researchers who plan to port their high-performance C++ codes to SYCL to get all the benefits of single-source programming.
\end{abstract}
%
% ---- Sections ----
%
\section{Introduction}
The financial industry solves many computationally intensive problems requiring high-performance computing. Over time some financial problems have become traditional workload benchmarks. The Black-Scholes option pricing problem \cite{black2019pricing} is one of the most widespread among them due to several reasons. Firstly, it is one of the basic elements of financial market analysis and, therefore, has a great practical interest. Secondly, ease of understanding and implementation, combined with high computational requirements, make the Black-Scholes model popular for learning the basics of optimization and performance testing on different architectures, including various accelerators such as GPU, Xeon Phi, and others \cite{meyerov2014performance,grauer2013accelerating,smelyanskiy2012analysis,podlozhnyuk2007black,pharr2005gpu}. In this paper, we consider the Black-Scholes formula for a fair price of a European call option and employ several optimization techniques to improve performance on CPUs and GPUs. Our main motivation is to understand is it possible to achieve reasonable performance on Intel CPUs and GPUs using a new oneAPI programming model. 

oneAPI \cite{oneapi} is a new technology of single-source heterogeneous programming, introduced by Intel in 2020. Just like OpenCL \cite{opencl}, OpenACC \cite{openacc}, Kokkos \cite{edwards2013kokkos}, Alpaka \cite{zenker2016alpaka} and many others, it is a software framework that allows implementing a single code for various hardware. With a growing diversity of architectures and their wide distribution, including among supercomputer systems, such frameworks are of increasing interest for developing high-performance applications in various problem areas. oneAPI provides a large set of tools for efficient development and performance analysis and optimization on CPUs, GPUs, FPGAs, and other devices. oneAPI includes the Data Parallel C++ (DPC++) programming language \cite{reinders2021data} based on the SYCL model \cite{reyes2016sycl,sycl}. DPC++ and SYCL are almost the same languages, so both designations will be used in the same sense in the rest of the paper.

The opportunity to write a single portable code that can be compiled and run on a variety of computing devices of various architectures and vendors is an important advantage of DPC++. This feature could allow us to develop a single portable code instead of a family of codes optimized for each computer architecture. However, whether \textit{performance portability} is possible to achieve is still the open question.

In this paper, we address this important question by calculating the fair prices of a set of European call options. Compared to the previous paper \cite{meyerov2014performance} we analyze which optimization techniques to the C++ code parallelized with OpenMP improve performance on modern CPUs. We consider this implementation as a baseline for CPUs. After that, we show how to port the code to DPC++/SYCL and achieve reasonable performance not only on Intel CPUs but also on Intel GPUs. We highlight which optimization techniques are beneficial for this memory-bound application, and hope that our experience will be useful to other developers.

The paper is organized as follows. Section~\ref{sec:model} provides the Black-Scholes formula and the corresponding algorithm. Test infrastructure and some launch parameters used in computational experiments are described in Section~\ref{sec:test_infrastructure}. The OpenMP implementation on C++ and its step-by-step optimization are presented in Section~\ref{sec:cpp_implementation}. In Section~\ref{sec:dpcpp_implementation} we propose the DPC++ implementation of the Black-Scholes formula and optimize it on Intel CPUs and GPUs. Section~\ref{sec:conclusion} summarizes the paper.

\section{Black-Scholes Option Pricing}\label{sec:model}
We consider the European share call option, which is the simplest variant of an option and describes the following contract between two parties. The first party promises to sell the second one a share at a fixed strike price $K$ at some point in time $T$ after the contract conclusion, and the second party pays the first party an amount $C$ for this right. After the maturity of $T$, the second party can decide to buy the share or not, it depends on its current price $S_T$: if $S_T-K-C>0$ then it is profitable. The question is how to define the fair option price $C$, which is a balance of gains and losses for each party. It can be determined by the Black-Scholes formula for European call option price:
\begin{eqnarray}
\label{eq:bs_formula_1}
C = S_0F(d_1)-Ke^{-rT}F(d_2) \\
\label{eq:bs_formula_2}
d_1 = \frac{\ln{\frac{S_0}{K}} + \left(r + \frac{\sigma^2}{2}\right)T}{\sigma\sqrt{T}} \\
\label{eq:bs_formula_3}
d_2 = \frac{\ln{\frac{S_0}{K}} + \left(r - \frac{\sigma^2}{2}\right)T}{\sigma\sqrt{T}},
\end{eqnarray}
where $S_0>0$ is a starting price of a share, $r$ is an interest rate (describes how many times the price of bonds in the financial market will increase), $\sigma$ is volatility (a coefficient of risk assessment), $K$ is a strike price, $T$ is maturity, $C$ is an option price, and $F(x)$ is a cumulative normal distribution function. We do not go into detail about key assumptions and ideas that lead to the Black-Scholes formula, this is fully described in \cite{black2019pricing,meyerov2014performance}. In what follows, we will assume that $r$ and $\sigma$ are fixed, while $S_0$, $K$ and $T$ depend on the current state of the financial market. The main intention is to calculate the optimal option price $C$ based on the given input parameters.

In practice, prices are calculated for a huge amount of different options for many specific market conditions. The performance of financial calculations significantly influences the decision-making speed, so reducing computational time for a large set of options is very important. Below we present the Black-Scholes formula implementation on C++ and DPC++ programming languages and optimize it step by step for CPUs and GPUs. We demonstrate that although the formula looks very simple to implement, it is a good testing ground for employing and analyzing a wide range of ideas related to performance optimization, especially when using devices of different architectures. 

\section{Test Infrastructure}\label{sec:test_infrastructure}
Experiments were performed at the following infrastructures:
\begin{itemize}
    \item A node of the Endeavour supercomputer with 2x Intel Xeon Platinum 8260L (Cascade Lake, 24 cores each, 2400~MHz), 192 GB RAM, RedHat 4.8.5, Intel C++ Compiler Classic, and Intel DPC++ Compiler from the Intel OneAPI Toolkit 2022.1 (experiments on CPUs).
    \item A node of the Intel DevCloud platform with Intel Iris Xe MAX (96 EU, 1650~MHz, 4~GB RAM, 68~GB/sec of memory bandwidth), Intel Core i9 10920X, Ubuntu 18.04 LTS, Intel DPC++ Compiler from the Intel OneAPI Toolkit 2022.1 (experiments on GPUs).
\end{itemize}
According to peak performance and memory bandwidth, the designated GPU loses to the CPU, so we expect a worse performance on the GPU compared to the CPU. It is confirmed in practice. However, optimizations for such GPUs are of interest due to the prospect of new server GPUs being released by Intel.

The whole set of options is processed in 5 batches of 240,000,000 elements (each batch takes $\sim$3.8~GB). The minimum time for all batches is chosen as the resulting execution time. It is relevant for DPC++ experiments, as this way the DPC++ kernel compilation time, which increases the processing time of the first batch, is excluded from our measurements.

\section{C++ Implementation and Optimization}\label{sec:cpp_implementation}
\subsection{Baseline}\label{sec:cpp_baseline}
Even though the formulas \ref{eq:bs_formula_1}-\ref{eq:bs_formula_3} look quite simple, everything is not as elementary as it seems. First, we need to decide on the choice of a relevant data type and memory structure. The nature of the problem implies that it does not require high computational accuracy, so we can use single precision ('float' type) or even half precision (the last one is the subject of further research). It is worth noting that mixing 'double' and 'float' types is a typical mistake of many programmers: usage of double precision literals and inappropriate mathematical functions can significantly increase execution time. In its turn, careful handling of data types has only a positive effect on performance.

Employing a cache-friendly memory access pattern is one of the key performance optimization techniques. Applied to the Black-Scholes formula, we have two options to organize data: Array of Structures (AoS) and Structure of Arrays (SoA). In the AoS pattern case, all data related to the first option is placed into memory continued by data from the second option, and so on. In the SoA case, the data is packed in multiple arrays, and each array is responsible for storing a single parameter (initial price of a share, strike price, maturity, option price). More details about the advantages and disadvantages of each memory pattern for the option pricing problem are presented in \cite{meyerov2014performance}. Here we only want to point out that while SoA has given quite a significant speedup over AoS for the Black-Scholes code several years ago \cite{meyerov2014performance,smelyanskiy2012analysis}, now the difference has been disappeared. The AoS pattern works even a little faster than SoA on hardware described in Section~\ref{sec:test_infrastructure}. However, we need to develop code that is generally efficient across different architectures. Although the gap between SoA and AoS on the modern infrastructure does not exceed 1\%, on older infrastructures SoA gives a significant increase (up to 3 times) due to the better use of vector instructions \cite{meyerov2014performance}. Therefore, we still use the SoA pattern. 

The next issue that needs to be addressed is the choice of an appropriate mathematical library. Note that the cumulative normal distribution function ($F(x)$ in Eq.~\ref{eq:bs_formula_1}-\ref{eq:bs_formula_3}) is of particular interest. For example, it is implemented as \texttt{cdfnorm()} function in the Intel Math Library of Intel C++ Compiler Classic. A more flexible and cross-platform solution is to use mathematical functions from the C++ language standard, which does not contain the normal distribution function but implements an error function and a similar complementary error function. In future computations we offer to use the next formula (Eq.~\ref{eq:cdfnorm_as_erf}). Note that it can work even faster than cdfnorm() in some implementations due to simpler compute approximation.
\begin{eqnarray}
\label{eq:cdfnorm_as_erf}
F(x)=0.5+0.5\erf{\frac{x}{\sqrt{2}}}
\end{eqnarray}

Considering all of the above, the baseline implementation of the Black-Scholes formula is presented in Listing~\ref{lst:cpp_baseline}. Its execution time on the CPU presented in Section~\ref{sec:test_infrastructure} was 3.062~sec (Table~\ref{tab:cpp_versions}).

\begin{lstlisting}[label=lst:cpp_baseline, caption=Baseline C++ implementation of the Black-Scholes formula.]
const float sig = 0.2f;    // volatility (0.2 -> 20%)
const float r = 0.05f;     // interest rate (0.05 -> 5%)
// the following constants are used
// to initialize pT, pS0 and pK arrays
const float T = 3.0f;      // maturity (3 -> 3 years)
const float S0 = 100.0f;   // initial stock price
const float K = 100.0f;    // strike price

void GetOptionPrices (float *pT, float *pK, float *pS0,
  float *pC, int N) {
  for (int i = 0; i < N; i++)
  {
    float d1 = (std::log(pS0[i]/pK[i]) + (r + 0.5f*sig*sig)*
               pT[i]) / (sig * std::sqrt(pT[i]));
    float d2 = (std::log(pS0[i]/pK[i]) + (r - 0.5f*sig*sig)*
               pT[i]) / (sig * std::sqrt(pT[i]));
    float erf1 = 0.5f + 0.5f*std::erf(d1 / std::sqrt(2.0f));
    float erf2 = 0.5f + 0.5f*std::erf(d2 / std::sqrt(2.0f));
    pC[i] = pS0[i]*erf1 - pK[i] * 
            std::exp((-1.0f)*r*pT[i])*erf2;
  }
}
\end{lstlisting}

\subsection{Loop Vectorization}\label{sec:cpp_vectorization}
Vectorization is one of the most effective optimization techniques. As to the Black-Scholes code, the absence of data dependencies between loop iterations and a suitable memory pattern are expected to provide high performance of vector calculations, which is limited by the vector register length ($\sim$8x for 256-bit registers and $\sim$16x for 512-bit registers). Modern optimizing compilers are effective enough when it comes to vectorization; the compiler report gives detailed information on whether vector calculations are involved or not and why, as well as what speedup is expected from vectorization. As a rule, for successful vectorization, a programmer needs to specify the absence of data dependencies explicitly because it is often impossible for the compiler to prove that computations are independent even if it is true. Moreover, we have noticed that for the baseline Black-Scholes code (Section~\ref{sec:cpp_baseline}) and the test infrastructure we used (Section~\ref{sec:test_infrastructure}) the compiler has generated two versions of the code at once, one vectorized and one not. In this case, the decision on the possibility of using vectorized code is deferred until the program is run and much more information is available. Often the vectorized version is chosen at runtime. Thus, we just had to define a set of AVX512 instructions (\textbf{-xHost}, \textbf{-qopt-zmm-usage=high} compiler keys) instead of default SSE instructions to better utilize CPU vector units. As a result, execution time has been reduced to 1.155~sec ($\sim$2.5x speedup).

\subsection{Precision Reduction}
We said above that the specificity of this particular financial problem allows us to involve inaccurate but fast calculations of mathematical functions. Accordingly, we used the next options of the Intel compiler:
\begin{itemize}
    \item \textbf{fimf-precision=low} tells the compiler to use mathematical function implementations with only 11 accurate bits in mantissa instead of 24 bits available in single precision.
    \item \textbf{fimf-domain-exclusion=31} allows the compiler to exclude special values in mathematical functions, such as infinity, NaN, or extreme values of function arguments.
\end{itemize}

This technique has given a 2x speedup compared to the previous version (0.527~sec, see Table~\ref{tab:cpp_versions}). Need to mark, we got a different and less accurate result (20.927 instead of a reference 20.924 for the start parameters presented in Listing~\ref{lst:cpp_baseline}), but the difference does not occur until the 5th decimal digit, which is acceptable for financial models. However, the precision controls should be used with care and properly analyzed for the accuracy of the result.

\subsection{Parallelism}\label{sec:cpp_parallelism}
In order to use the maximum potential of a CPU, it is expedient to distribute the computational load between its physical cores. We used the OpenMP model; in the case of the Black-Scholes code, it was enough to add \textbf{omp parallel for} pragma before the main loop and specify \textbf{-qopenmp} compiler option. The processor on which the experiments were carried out has 48 physical cores (2 CPUs with 24 cores each) and supports hyper-threading technology. The thread affinity mask is set to 'compact'. The maximal speedup of the parallel code was achieved on 48 threads, hyper-threading did not impact the performance. The best execution time was 0.053~sec, that is 10 times faster (Table~\ref{tab:cpp_versions}). The reason for which we do not observe better speedup is limited memory bandwidth, and we can increase performance through more efficient memory management (see Section~\ref{sec:cpp_numa} for more details).

\subsection{NUMA-Friendly Memory Allocation}\label{sec:cpp_numa}
The heterogeneous memory structure is a characteristic of many modern CPUs. Non-Uniform Memory Access (NUMA) implies that a processor can access its memory faster than the memory of another processor on the same cluster node. We use a two-socket system, and in the worst case, we can get 50\% remote (and inefficient) access because all memory is allocated on only one processor. To avoid this, it is enough to allocate memory in a NUMA-friendly manner. According to the first-touch policy \cite{hager2010introduction}, we should care about the appreciate initialization of a current option batch in a parallel loop before the main calculations, as shown in Listing~\ref{lst:cpp_numa}. This trick gave the 2x speedup, the execution time has become final 0.022~sec (Table~\ref{tab:cpp_versions}).

\begin{lstlisting}[label=lst:cpp_numa, caption=NUMA-friendly initialization.]
#pragma omp parallel for simd
for (int i = 0; i < N; i++) {
  pT[i] = T;
  pS0[i] = S0;
  pK[i] = K;
  pC[i] = 0;
} 
\end{lstlisting}

\begin{figure}[ht]
\center\includegraphics[width=0.98\linewidth]{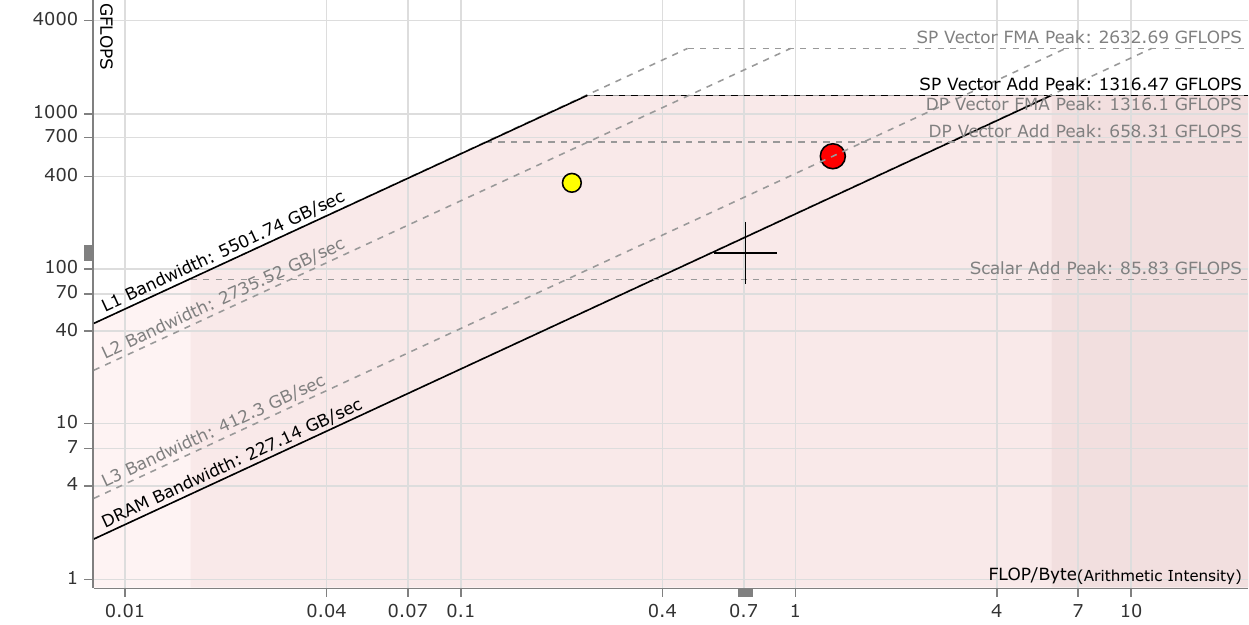}
\caption[]{The Roofline model of the optimized Black-Scholes C++ code on the CPU. The diagram is plotted by Intel Advisor from OneAPI Toolkit 2022.1. The red point denotes the main Black-Scholes computational loop.}
\label{fig:cpu_roofline}
\end{figure}

Nevertheless, we have not reached the maximal speedup in parallel mode (20x instead of maximal 48x). The Roofline model \cite{marques2017performance} shown in Figure~\ref{fig:cpu_roofline} indicates that performance is limited by L3 cache bandwidth. In this case, the most common technique is to increase data locality, for example, by using cache blocking techniques, but it does not apply to the Black-Scholes code. Likely, we have reached the execution time limit.

All optimization stages of the C++ Black-Scholes code is presented in Table~\ref{tab:cpp_versions}. We can see that the total speedup through all optimizations and parallelization compared to the baseline serial code is $\sim$140x.
\begin{table}[ht]
    \caption[]{Execution time of each C++ optimization stage.}
    \centering
    \begin{tabular}{|l|c|}
        \hline
        ~\textbf{Version}~&~\textbf{Time, sec}~\\
        \hline
        ~Baseline ~&~ 3.062 ~\\
        ~Loop Vectorization ~&~ 1.155 ~\\
        ~Precision Reduction ~&~ 0.527 ~\\
        ~Parallelism ~&~ 0.053 ~\\
        ~NUMA-Friendly Memory Allocation ~&~ \textit{0.022} ~\\
        \hline
    \end{tabular}
    \label{tab:cpp_versions}
\end{table}

\section{Porting to DPC++}\label{sec:dpcpp_implementation}
\subsection{Baseline and Optimizations on CPUs}\label{sec:dpcpp_cpu_baseline}
The effective implementation of any algorithm on various types of accelerators implies its decomposition to a sufficiently large number of small subtasks that can be performed in parallel. In the baseline Black-Scholes formula implementation on DPC++, we choose a single option price calculation as a work item and run all of them in parallel on a device (see Listing~\ref{lst:dpc_baseline_kernel}). We use SYCL built-in mathematical functions.

\begin{lstlisting}[label=lst:dpc_baseline_kernel, caption=Baseline DPC++ kernel implementation of the Black-Scholes formula and its parallel launch.]
handler.parallel_for(sycl::range<1>(N), [=](sycl::id<1> i) {
  float d1 = (sycl::log(pS0[i] / pK[i]) + (r + sig2*0.5f) *
              pT[i]) / (sig * std::sqrt(pT[i]));
  float d2 = (sycl::log(pS0[i] / pK[i]) + (r - sig2*0.5f) *
              pT[i]) / (sig * std::sqrt(pT[i]));
  float erf1 = 0.5f + 0.5f * sycl::erf(d1 * invsqrt2);
  float erf2 = 0.5f + 0.5f * sycl::erf(d2 * invsqrt2);
  pC[i] = pS0[i] * erf1 - pK[i] * sycl::exp((-1.0f) * r *
          pT[i]) * erf2;
});
\end{lstlisting}

When writing high-performance code for graphics accelerators, we should consider specifics of their memory management. Discrete GPUs require a smart data transfer policy between host and device. DPC++ provides different ways to do it, explicit or implicit. In the baseline Black-Scholes code we consider a convenient Buffers\&Accessors strategy \cite{reinders2021data} (Listing~\ref{lst:buffers_accessors}). More details about the various memory management strategies are presented in Section~\ref{sec:usm}.

\begin{lstlisting}[label=lst:buffers_accessors, caption=Buffers\&Accessors data management.]
void GetOptionPrices(float *pT, float *pK, float *pS0,
  float *pC, int N) {
  // declaration of buffers
  sycl::buffer<float, 1> pTbuf(pT, sycl::range<1>(N));
  sycl::buffer<float, 1> pKbuf(pK, sycl::range<1>(N));
  sycl::buffer<float, 1> pS0buf(pS0, sycl::range<1>(N));
  sycl::buffer<float, 1> pCbuf(pC, sycl::range<1>(N));
  
  queue.submit([&](auto& handler) {
    // declaration of accessors
    sycl::accessor pT(pTbuf, handler, sycl::read_only);
    sycl::accessor pK(pKbuf, handler, sycl::read_only);
    sycl::accessor pS0(pS0buf, handler, sycl::read_only);
    sycl::accessor pC(pCbuf, handler, sycl::write_only);
    // here the BS computational kernel should be
  }).wait();
  
  // data copying from device to host
  sycl::host_accessor pCacc(pCbuf);
  // here we need to process the result
}
\end{lstlisting}

As to DPC++ on CPUs, the concept of SYCL let us not think about vectorization and parallelism. However, we need to take into account non-uniform memory access to avoid problems described in Section~\ref{sec:cpp_numa}, otherwise, we get a time similar to that presented in Section~\ref{sec:cpp_parallelism} (Table~\ref{tab:dpcpp_versions_cpu}, Baseline). We can allocate memory in a NUMA-friendly manner by, for example, initializing parameter arrays in the separate DPC++ kernel (Listing~\ref{lst:dpc_numa_kernel}). 

\begin{lstlisting}[label=lst:dpc_numa_kernel, caption=Initialization in DPC++ kernel.]
handler.parallel_for(sycl::range<1>(N), [=](sycl::id<1> i) {
  pT[i] = T;
  pS0[i] = S0;
  pK[i] = K;
  pC[i] = 0;
});
\end{lstlisting}

Nevertheless, when we just added this piece of code, the execution time was reduced to 0.035~sec only instead of the expected 0.022~sec observed in the case of optimized C++ code (Section~\ref{sec:cpp_numa}). The issue is related to the TBB library which is used by DPC++ in this scenario. TBB threads distribute work dynamically without regard to NUMA. As a result, we have some percentage of remote memory access even after the appreciate initialization. To avoid this, the \textbf{DPCPP\_CPU\_PLACES} environment variable with the \textbf{numa\_domains} value should be set. In this case, dynamic scheduling of work within the TBB-based runtime occurs within each processor separately, which removes access to remote memory. After that, we got final 0.024~sec, which is only 10\% slower than the optimized C++ implementation (Table~\ref{tab:dpcpp_versions_cpu}). Taking into account that DPC++ code can be run on CPUs and GPUs, we consider 10\% of performance to be a reasonable price to pay for portability.

\begin{table}[ht]
    \caption[]{Execution time of each DPC++ optimization stage on CPUs.}
    \centering
    \begin{tabular}{|l|c|}
        \hline
        ~\textbf{Version}~&~\textbf{Time, sec}~\\
        \hline
        ~Baseline ~&~ 0.057 ~\\
        ~Initialization in a separate kernel ~&~ 0.035 ~\\
        ~NUMA-Friendly (DPCPP\_CPU\_PLACES) ~&~ \textit{0.024} ~\\
        \hline
        ~Optimized C++ code ~&~ 0.022 ~\\
        \hline
    \end{tabular}
    \label{tab:dpcpp_versions_cpu}
\end{table}

\subsection{Experiments on GPUs}\label{sec:dpcpp_gpu_baseline}
\begin{figure}[ht]
\center\includegraphics[width=0.98\linewidth]{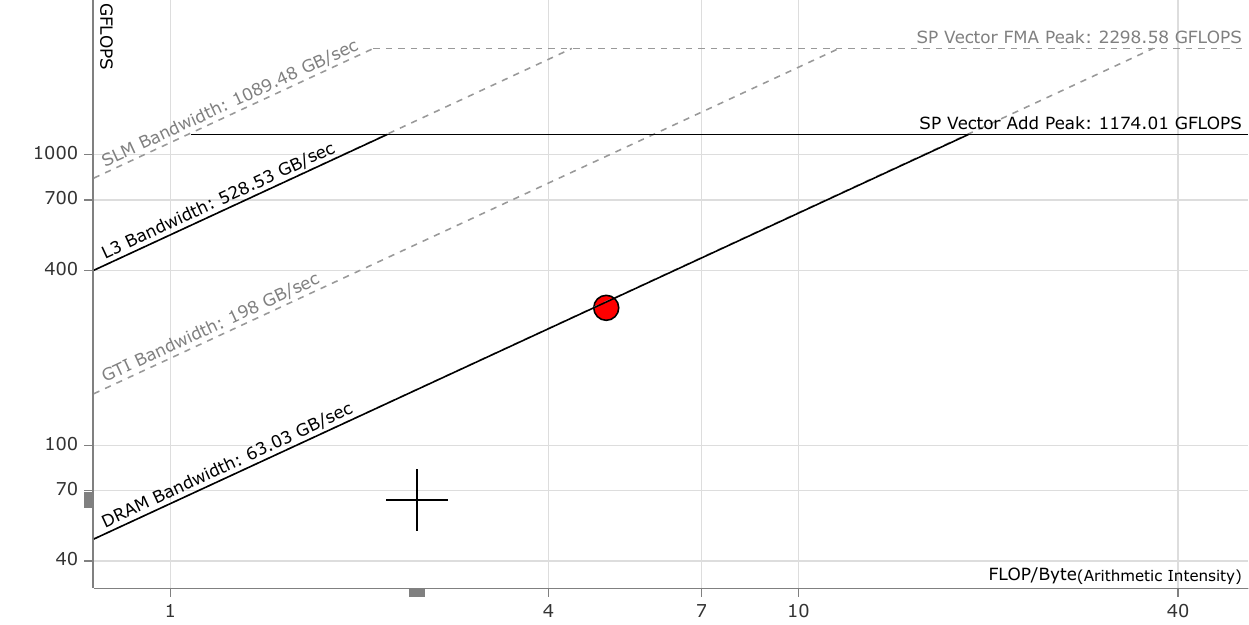}
\caption[]{The Roofline model of the Black-Scholes DPC++ code on the GPU. The diagram is plotted by Intel Advisor from OneAPI Toolkit 2022.1. The red point denotes the main computational kernel.}
\label{fig:gpu_roofline}
\end{figure}

The great advantage of the DPC++ model is that a DPC++ code can be run on many devices: CPUs, GPUs, and others. However, it does not mean that the same code will work optimally on all the devices, and we still need to take into account the device specifics when developing code. But it is worth noting that the GPU-optimized code often shows high performance on CPUs as well, so it makes sense to develop and optimize for GPUs first. Some common GPU-specific optimizations are presented in Section~\ref{sec:gpu_opt_tricks}, but all of them have not given any positive result for the benchmark at stake. The reason for it is the DRAM memory bandwidth limitations of the GPU (Figure~\ref{fig:gpu_roofline}).

The DPC++ baseline shows 0.062~sec of the kernel execution time, and this is the best result we got. It is worth saying that this is a fairly good result, corresponding to the characteristics of the device. Taking into account the characteristics of the memory bandwidth of this GPU and the fact that the benchmark is memory-bound, we expected the kernel execution time to be at least 0.056~sec (3.8~GB of batch size divided by 68~GB/sec of memory bandwidth).

The initialization kernel takes similar to the main kernel time -- 0.061~sec. However, we have noticed that the overhead of memory transfer and other operations is highly large -- more than 2~sec (Table~\ref{tab:dpcpp_versions_gpu}). In the next section, we demonstrate how to reduce these additional costs.

\subsection{Memory Management}\label{sec:usm}
As the experiments in Section~\ref{sec:dpcpp_gpu_baseline} show, it takes a lot of time to synchronize and copy data from host to GPU and back. In this section, we attempt to reduce overhead by using different memory management strategies provided by DPC++. The code presented in Section~\ref{sec:dpcpp_cpu_baseline} is based on the Buffers\&Accessors model, whereby we create a buffer object for each parameter array and use special accessor objects to read and write buffers. Based on buffers and accessors, a decision to copy data from host to device and back is made independently of a programmer. However, we can influence this by specifying at what point in time we need the data we are interested in on a particular device by creating a buffer or accessor object in the appropriate place in a DPC++ program. For example, in Listing \ref{sec:dpcpp_cpu_baseline}, the shared buffers for both the initialization and the option pricing kernels, as well as the absence of appropriate accessors between them, ensure that data is not transferred from device to host between these kernels.

It is worth paying attention that given that we process options in batches, we can allocate memory for only one batch and create the buffer objects once when the program starts. This technique had a good effect -- the overhead costs were reduced by 3 times (Table~\ref{tab:dpcpp_versions_gpu}).

An alternative approach to using buffers and accessors for data transfers relies on Unified Shared Memory (USM), which is introduced in DPC++ \cite{reinders2021data}. USM provides explicit and implicit ways to copy data from host to device and vice versa. The explicit way assumes direct APIs to move data between device and host. The implicit approach is based on shared memory support implemented in software and/or hardware, so users need just to allocate a shared memory region that will be accessible across all the devices automatically. In contrast to accessors, a user has to manage all the data dependencies manually while using USM, but at the same time, shared memory is much easier to adopt for "highly C++" codes. It is worth mentioning that the performance of USM data transfers is strongly dependent on its particular implementation and may differ for various devices. We review both implicit and explicit approaches in relation to the Black-Scholes formula and the infrastructure under consideration.

Following the implicit USM model, we create data pointers using the \linebreak\texttt{malloc\_shared()} function and then simply access them inside a kernel, with any necessary data movement done automatically (Listing~\ref{lst:usm_implicit}). In our case, this strategy slightly reduces data transfer time, but apparently requires additional synchronizations between the host and device when calling \texttt{malloc\_shared()}, which still does not matter if memory is allocated at once and common for all batches (Table~\ref{tab:dpcpp_versions_gpu}).

\begin{lstlisting}[label=lst:usm_implicit, caption=Implicit USM data movement.]
void GetOptionPrices(int N) {
  // shared memory allocation
  float *pT = sycl::malloc_shared<float>(N, queue);
  float *pS0 = sycl::malloc_shared<float>(N, queue);
  float *pK = sycl::malloc_shared<float>(N, queue);
  float *pC = sycl::malloc_shared<float>(N, queue);

  // here we need to run the initialization kernel
  // and the option price calculation kernel
  // and process the result pC
  
  // freeing memory
  sycl::free(pT); sycl::free(pS0);
  sycl::free(pK); sycl::free(pC);
}
\end{lstlisting}

It is worth noting that the resulting performance of implicit data movement in USM strongly depends on its internal architecture-specific implementation. Perhaps on newer hardware and software, the \texttt{malloc\_shared} approach will provide more benefits. As for the current infrastructure, we suspect that some data arrays are copied from host to device and back again a few extra times. Therefore, we decided to allocate memory directly on the device and explicitly transfer only the result array \texttt{pC} to the host under the explicit USM approach. The corresponding implementation shown in Listing~\ref{lst:usm_explicit} gave the best result (0.37~sec overhead, Table~\ref{tab:dpcpp_versions_gpu}).

\begin{lstlisting}[label=lst:usm_explicit, caption=Explicit USM data movement.]
void GetOptionPrices(int N) {
  // memory allocation on host to keep result 
  float *pCHost = new float[N];  
  // memory allocation on device
  float *pT = sycl::malloc_device<float>(N, queue);
  float *pS0 = sycl::malloc_device<float>(N, queue);
  float *pK = sycl::malloc_device<float>(N, queue);
  float *pC = sycl::malloc_device<float>(N, queue);

  // here we need to run the initialization kernel
  // and the option price calculation kernel
  
  // data copying from device to host
  queue.memcpy(pCHost, pC, N * sizeof(float)).wait();
  // here we need to process the result pCHost
  // and free all allocated memory
}
\end{lstlisting}

\begin{table}[ht]
    \caption[]{Different DPC++ memory management models: kernel execution time and overhead time on the GPU. 'One batch' means that we allocate memory for each option batch and include its time into measurements; 'multiple batches' means that we allocate memory that is shared for all batches at once, and do not include this into measurements. The table shows the time of one batch processing.}
    \centering
    \begin{tabular}{|l|c|c|c|c|}
        \hline
        %~\textbf{Version} ~&~ \textbf{Main kernel, sec} ~&~ \textbf{Initialization kernel, sec} ~&~ \textbf{Overhead, sec} ~&~ \textbf{Total, sec}~\\
        ~\textbf{Version} ~&~ \textbf{Main} ~&~ \textbf{Initiali-} ~&~ \textbf{Over-} ~&~ \textbf{Total,} ~\\
        ~&~ \textbf{kernel,} ~&~ \textbf{zation} ~&~ \textbf{head,} ~&~ \textbf{sec} ~\\
        ~&~ \textbf{sec} ~&~ \textbf{kernel, sec}  ~&~ \textbf{sec} ~&~ ~\\
        \hline
        ~Buffers\&Accessors, one batch ~&~ \multirow{6}{*}{0.062} ~&~ \multirow{6}{*}{0.061} ~&~ 2.367 ~&~ 2.551 ~ \\
        ~Buffers\&Accessors, multiple batches ~&~ ~&~ ~&~ 0.748 ~&~ 0.932 ~ \\
        ~USM, implicit, one batch ~&~ ~&~ ~&~ 2.987 ~&~ 3.110 ~ \\
        ~USM, implicit, multiple batches ~&~ ~&~ ~&~ 0.658 ~&~ 0.781 ~ \\
        ~USM, explicit, one batch ~&~ ~&~ ~&~ 1.158 ~&~ 1.281 ~ \\
        ~USM, explicit, multiple batches ~&~ ~&~ ~&~ \textit{0.369} ~&~ \textit{0.492}~ \\
        \hline
    \end{tabular}
    \label{tab:dpcpp_versions_gpu}
\end{table}

\subsection{Several Common Optimization Tricks}\label{sec:gpu_opt_tricks}
In this subsection, we look at important optimization techniques that can provide significant performance gains on GPUs. Note that in our particular case, they did not work for various reasons. In several cases, the compiler and runtime did a great job without our help. In other cases, the nature of the task, where memory access operations are a significant part of the work compared to calculations, did not allow for achieving additional speedup. In the meantime, it seemed important to us to demonstrate these techniques for performance optimization, as they may be useful to researchers and developers in other applications.

In the baseline DPC++ implementation, we start $N$ work items for every option. The initial one-dimensional range is divided into work-groups automatically by the DPC++ compiler and runtime. However, a programmer can vary a range, work-group or sub-group size, and here the potential for optimization lies. The trick presented in Listing~\ref{lst:groups} contributes to the solution of possible optimization problems such as underutilization of GPU cores, the overhead of maintaining a large number of threads, and suboptimal work-group size. The example demonstrates the variation in the total number of threads and the group size. It is worth paying attention that in the internal kernel loop it is necessary to walk through the iterations with a step equal to the number of threads to avoid memory bank conflicts. As to the Black-Scholes code, we checked different combinations of thread count and work-group size. Experiments on GPUs have shown that the optimal values are the maximal thread count and group size, that is, the default configuration used by the DPC++ compiler and runtime is the most beneficial.
\begin{minipage}{\linewidth}
\begin{lstlisting}[label=lst:groups, caption=An example of varying the number of threads and group size.]
handler.parallel_for(sycl::nd_range<1>(THREAD_COUNT,
  GROUP_SIZE), [=](sycl::nd_item<1> item) {
  for (int i=item.get_global_id(0); i<N; i+=THREAD_COUNT) {
    // the i-th element processing
  }});
\end{lstlisting}
\end{minipage}

Another advantageous technique is to use vector data types \cite{reinders2021data}. In this way, we explicitly state that some elements are arranged in a row in memory. It can positively affect the speed of loading and computing on GPUs as well as on CPUs. We have tried to use \textbf{sycl::float4} data type but, in our case, it has not impacted the performance.

It should also be said that we tried using the different memory patterns in an attempt to optimize it for GPUs. The AoS pattern did not improve execution time compared to SoA, but we have noticed that it is profitable to store read-only and write-only data separately to reduce the load on the performance-limiting DRAM.

\section{Conclusion}\label{sec:conclusion}
In this paper, we present a high-performance implementation of the Black-Scholes option pricing formula in the C++ and SYCL (DPC++) programming languages. We optimize it step by step and demonstrate how various optimization techniques improve performance for commonly applied Intel CPUs and GPUs. In particular, on CPUs, we employ and analyze vectorization, precision reduction, single-node parallelism, and optimal use of the NUMA architecture. For GPUs, we discuss device-friendly memory access patterns, relevant data management, and employing vector data types. Experiments showed that on CPUs, the optimized DPC++ implementation was only 10\% slower than the highly optimized OpenMP implementation, while it was possible to run it also on GPUs and observe the expected performance based on the device computational capabilities. We demonstrated that the performance was limited by memory bandwidth both on CPU and GPU. 

We hope that our results will be useful to other researchers who are planning to port their codes from C++ to SYCL (DPC++). In this regard, we discussed not only those optimization techniques that improved performance but also those that did not speed up our application. We believe that these techniques can work in other codes. They also can help on upcoming devices with significantly larger memory bandwidth. In general, the emergence and development of SYCL ideas within the oneAPI model looks like a promising direction in modern High Performance Computing.

\section*{Acknowledgements}\label{sec:acknowledgements}
{
The work is supported by the oneAPI Center of Excellence program and by the Ministry of Science and Higher Education of the Russian Federation, project no. 0729-2020-0055.
}

%\begin{thebibliography}{6}
%
%
%\bibitem {smit:wat}
%Smith, T.F., Waterman, M.S.: Identification of common molecular subsequences.
%J. Mol. Biol. 147, 195?197 (1981). \url{doi:10.1016/0022-2836(81)90087-5}
%
%\end{thebibliography}
%
\bibliographystyle{spmpsci_unsrt}
\bibliography{literature}

\begin{thebibliography}{10}
\providecommand{\url}[1]{{#1}}
\providecommand{\urlprefix}{URL }
\expandafter\ifx\csname urlstyle\endcsname\relax
  \providecommand{\doi}[1]{DOI~\discretionary{}{}{}#1}\else
  \providecommand{\doi}{DOI~\discretionary{}{}{}\begingroup
  \urlstyle{rm}\Url}\fi

\bibitem{black2019pricing}
Black, F., Scholes, M.: The pricing of options and corporate liabilities.
\newblock In: World Scientific Reference on Contingent Claims Analysis in
  Corporate Finance: Volume 1: Foundations of CCA and Equity Valuation, pp.
  3--21. World Scientific (2019)

\bibitem{meyerov2014performance}
Meyerov, I., Sysoyev, A., Astafiev, N., Burylov, I.: Performance optimization
  of black-scholes pricing.
\newblock In: High Performance Parallelism Pearls: Multicore and Many-core
  Programming Approaches, pp. 319--340. Springer, Cham (2014)

\bibitem{grauer2013accelerating}
Grauer-Gray, S., Killian, W., Searles, R., Cavazos, J.: Accelerating financial
  applications on the gpu.
\newblock In: Proceedings of the 6th Workshop on General Purpose Processor
  Using Graphics Processing Units, pp. 127--136 (2013)

\bibitem{smelyanskiy2012analysis}
Smelyanskiy, M., Sewall, J., Kalamkar, D.D., Satish, N., Dubey, P., Astafiev,
  N., Burylov, I., Nikolaev, A., Maidanov, S., Li, S., et~al.: Analysis and
  optimization of financial analytics benchmark on modern multi-and many-core
  ia-based architectures.
\newblock In: 2012 SC Companion: High Performance Computing, Networking Storage
  and Analysis, pp. 1154--1162. IEEE (2012)

\bibitem{podlozhnyuk2007black}
Podlozhnyuk, V.: Black-scholes option pricing (2007)

\bibitem{pharr2005gpu}
Pharr, M., Fernando, R.: GPU Gems 2: Programming techniques for
  high-performance graphics and general-purpose computation (gpu gems).
\newblock Addison-Wesley Professional (2005)

\bibitem{oneapi}
{oneAPI: A New Era of Heterogeneous Computing}.
\newblock
  \url{https://www.intel.com/content/www/us/en/developer/tools/oneapi/overview.html}

\bibitem{opencl}
{OpenCL: open standard for parallel programming of heterogeneous systems.}
\newblock \url{https://www.khronos.org/opencl/}

\bibitem{openacc}
{OpenACC}.
\newblock \url{https://www.openacc.org/}

\bibitem{edwards2013kokkos}
Edwards, H.C., Trott, C.R.: Kokkos: Enabling performance portability across
  manycore architectures.
\newblock In: 2013 Extreme Scaling Workshop (xsw 2013), pp. 18--24. IEEE (2013)

\bibitem{zenker2016alpaka}
Zenker, E., Worpitz, B., Widera, R., Huebl, A., Juckeland, G., Kn{\"u}pfer, A.,
  Nagel, W.E., Bussmann, M.: Alpaka--an abstraction library for parallel kernel
  acceleration.
\newblock In: 2016 IEEE International Parallel and Distributed Processing
  Symposium Workshops (IPDPSW), pp. 631--640. IEEE (2016)

\bibitem{reinders2021data}
Reinders, J., Ashbaugh, B., Brodman, J., Kinsner, M., Pennycook, J., Tian, X.:
  Data parallel C++: mastering DPC++ for programming of heterogeneous systems
  using C++ and SYCL.
\newblock Springer Nature (2021)

\bibitem{reyes2016sycl}
Reyes, R., Lom{\"u}ller, V.: Sycl: Single-source c++ accelerator programming.
\newblock In: Parallel Computing: On the Road to Exascale, pp. 673--682. IOS
  Press (2016)

\bibitem{sycl}
{SYCL}.
\newblock \url{https://www.khronos.org/sycl/}

\bibitem{hager2010introduction}
Hager, G., Wellein, G.: Introduction to high performance computing for
  scientists and engineers.
\newblock CRC Press (2010)

\bibitem{marques2017performance}
Marques, D., Duarte, H., Ilic, A., Sousa, L., Belenov, R., Thierry, P.,
  Matveev, Z.A.: Performance analysis with cache-aware roofline model in intel
  advisor.
\newblock In: 2017 International Conference on High Performance Computing \&
  Simulation (HPCS), pp. 898--907. IEEE (2017)

\end{thebibliography}

\end{document}